\documentstyle[preprint,aps]{revtex}
\input epsf.tex
\def\DESepsf(#1 width #2){\epsfxsize=#2 \epsfbox{#1}}
\begin{document}
\preprint{\vbox{\hbox{OITS-593}\hbox{UCRHEP-T157}\hbox{hep-ph/9512398}}} 
\draft
\title{Intermediate scale as a
source of lepton flavor violation in SUSY SO(10)}
\author{N. G. Deshpande $^{\ast}
$, B. Dutta$^{\ast}
$ , and E. Keith
$^{\dagger} $ }
\address{$^{\ast} $ Institute of Theoretical Science, University of
Oregon, Eugene, OR 97403 \\$^{\dagger} $ Department of Physics, University of
California, Riverside, CA 92521}
\date{December, 1995}
\maketitle
\begin{abstract}
In supersymmetric SO(10) grand unified models,
we examine the lepton flavor violation process $\mu \rightarrow e\gamma$ from
having the SU(2)$_R\times $U(1)$_{B-L}$ gauge symmetry broken at an intermediate
scale $M_I$ below the SO(10) grand unification scale $M_G$. Even in the case
that supersymmetry is broken by universal soft terms introduced at the scale
$M_G$, we find  significant rates for $\mu \rightarrow e\gamma$ with $M_I
\sim 10^{12}$ GeV or less. These rates are further enhanced if the universal soft
terms appear at a scale greater than
$M_G$. 
\end{abstract}

It has recently been pointed out \cite{[LJ],[AS],[cia]} that significant lepton
flavor violation can arise in supersymmetric (SUSY) grand unified theories. The
origin of this flavor violation resides in the largeness of the top Yukawa
coupling and the assumption that supersymmetry is broken by flavor uniform soft
breaking terms communicated to the visible sector by gravity at a scale $M_X$.
Assuming that $M_X$ is the reduced Planck scale which is much greater than the
grand unification scale $M_G$, renormalization effects cause the third
generation multiplet of squarks and sleptons which belong to the same multiplet
as the top in the grand unified theory (GUT) to become lighter than those of the
first two generations. The slepton and the charged lepton mass matrices can no
longer be simultaneously diagonalized thus inducing lepton flavor violation
through a suppression of the GIM mechanism in the slepton sector. This effect is
more pronounced in SO(10) models than in SU(5) where the left-handed slepton
mass matrices remain degenerate. The evolution of soft terms from $M_X$ to $M_G$
causes these flavor violations, which disappear when $M_X=M_G$. Here,
we explore another class of theories which are SUSY SO(10) GUTs which break down
to an intermediate gauge group $G_I$ before being broken to the standard model
(SM) gauge group at the scale $M_I$. In this class of theories, even if
$M_X=M_G$, lepton flavor violation arises due to the effect of the third
generation neutrino Yukawa coupling on the evolution of the soft leptonic terms
from the grand unification scale to the intermediate scale. Depending on the
location of the intermediate scale
$M_I$ and the size of the top Yukawa coupling at $M_G$, these rates can be
within one order of magnitude of the current experimental limit. Our results
will also indicate that if
$M_X>M_G$ in SUSY SO(10) models with an intermediate scale, the predicted rates
of lepton violating processes are further enhanced. We will concentrate on the
decay
$\mu\rightarrow e\gamma$ as an example since experimentally it is likely to be
the most viable.

SO(10) \cite{[GFM]} has many outstanding virtues that recommend it as a group
for grand unification. Among them are: (1) all fermions are in a single
representation, (2) a possibility exists to understand the observed patterns of
fermion masses and mixing \cite{[SOTM]} due in large part to useful SO(10)
Clebsch factors, (3) small nonzero neutrino masses are generated through the
see-saw mechanism
\cite{[SEE],[BM93]} and, (4) a scenerio for baryogenesis is possible
\cite{[FY86]}. Further, SUSY is an attractive feature for GUTs because it
provides a solution to the fine-tuning problem in the Higgs potential and, as a
bonus, the third generation SM fermion masses are more easily accounted for
\cite{[DK]}. In addition, SUSY SO(10) can accomodate a simple mechanism to solve
the doublet-triplet splitting problem \cite{[DiWi]}. (SUSY) SO(10) also has the
interesting feature that it allows an intermediate scale $M_I$ before breaking
to the standard model.   In models where $M_I \sim 10^{10}-10^{12}$ GeV, one can
naturally  get a neutrino mass in the interesting range of $\sim 3-10$ eV, which
could serve as hot dark matter which may be needed to explain the observed large
scale structure formation of the universe \cite{[SS]}. In models without an
intermediate gauge symmetry, in principle one could  produce a tau-neutrino
Majorana mass that is much less than the GUT breaking scale as for example via
non-renormalizable operators involving Higgs in the SO(10) ${\overline 16}$
representation or a  small and carefully chosen  Yukawa coupling to a
${\overline 126}$ field. However,  this would suffer from the further problem of
abandoning
$b-\tau$ Yukawa coupling unification except possibly in the case of high
$\tan\beta$ \cite{[ML10]}. As we have verified, this problem is signifigantly
reduced for the case of an intermediate scale \cite{[MohaLee],[BM]}.  The window
$\sim 10^{10}-10^{12}$ GeV is also of the right size for a hypothetical
PQ-symmetry to be broken so as to solve the strong CP problem without creating
phenomenological or  cosmological problems \cite{[axion]}.    Models which allow
$M_I\sim 1$ TeV are also interesting since they would predict relatively light
new gauge fields, as for example SU(2)$_R$ charged gauge bosons $W_R$. The
thrust of this study will be on flavor non-conservation in the leptonic sector
for
$M_I<M_G$ and $M_X = M_G$ in SUSY SO(10). Towards this goal, we study one model
where
$M_I
\sim 10^{11}-10^{12} $ GeV range, one where $M_I$ can be as low as the TeV range
with left-right gauge symmetry being preserved in $G_I$, and one where $M_I$
could have any value between the TeV scale and the GUT breaking scale without
violating any bounds on gauge couplings obtained from Z-physics.

We will consider scenarios for which the intermediate gauge symmetry is
$G_I\equiv$SU(2)$_L\times$SU(2)$_R\times$U(1)$_{B-L}\times$SU(3)$_c$ although
the concept that we illustrate in this paper is applicable to any GUT breaking
scenario for which an intermediate gauge symmetry exists under which
right-handed neutrinos transform non-trivially. For the purpose of keeping the
calculation as simple as possible,  we will consider only the
situation of a low value of
$\tan{\beta}$. We will choose to use the arbitrary value $\tan\beta =2$ in our
examples.  (However, analagous to the dependance on
$\tan\beta$ in the case of lepton flavor violation generated by physics above
the GUT breaking scale \cite{[cia],[us]}, using larger values of
$\tan\beta$ tends to produce even greater values of lepton flavor
violation.)
$\tan\beta$ not being $\sim m_t/m_b$ requires two bidoublets, which transform as
$(2,2,0,1)$ under the gauge symmetry
$G_I$, to account for all the standard model fermion masses in a natural
fashion.
As a consequence the Yukawa couplings that give masses
to the tau lepton and bottom quark are small enough so that terms of order
$\lambda_b^2$ may be neglected in the RGEs. Below the scale $M_I$ only the
Yukawa coupling
$\lambda_t$ is relatively large, however above $M_I$ the top Yukawa coupling and
the tau-neutrino Yukawa coupling are both large. In fact, at the GUT scale these
two couplings are equal. The presence of this large tau-neutrino Yukawa coupling
above the intermediate breaking scale causes the third generation slepton masses
to be less than the other two generations, and hence the lepton falvor violation
gets generated. It is also possible to produce lepton flavor violation in the
minimal supersymmetric standard model (MSSM) with the introduction of an
arbitrary intermediate scale $v_R \sim 10^{12}$ GeV for Majorana masses as in
ref.\cite{[Y]} and $M_X$ again taken at the reduced Planck scale. We do not find
this approach theoretically well motivated, and it also suffers from the
technical shortcomings of the unification scale being taken at the reduced
Planck scale while the MSSM gauge couplings appearantly unify at a much lower
scale
$M_G$ and the scale of $v_R$ not being associated with any breaking of the gauge
symmetry.

The superpotential terms which will be responsible for giving the SM fermion
masses have the following form when $G_I$ is the gauge symmetry:
\begin{eqnarray} W_Y&=&{\bf \lambda_{Q_u}}{\bf Q_L}{\Phi_2}{\bf Q_R} +{\bf
\lambda_{L_\nu}}{\bf L_L}{\Phi_2}{\bf L_R}\nonumber \\ &+& {\bf
\lambda_{Q_d}}{\bf Q_L}{\bf \Phi_1}{\bf Q_R} +{\bf \lambda_{L_e}}{\bf
L_L}{\Phi_1}{\bf L_R}\, ,
\end{eqnarray} where all group and generation indicies have beeen suppressed,
and $Q_{L,R}$ and $L_{L,R}$ represent the quark and lepton superfields which
transform as doublets under SU(2)$_L$ or SU(2)$_R$ and $\Phi_1$ and $\Phi_2$ are
the two bidoublets. We have assumed that $\Phi_2$ contains the MSSM Higgs
doublet which gives masses to the up quarks and Dirac masses for the neutrinos.
$\Phi_1$ contains the doublet which gives masses to the down quarks and the
charged leptons. These Yukawa couplings run in $G_I$ as follows:
\begin{eqnarray} {\cal D}\ln\lambda^2_{Q_{uj}} &=& -\sum_i{c^{\left(\lambda_Q
\right)}_ig^2_i}+ \left(3+4\delta_{j3}\right)\lambda_{Q_t}^2+
\lambda_{\nu_\tau}^2,\\ {\cal D}\ln\lambda^2_{Q_{dj}} &=&
-\sum_i{c^{\left(\lambda_Q \right)}_ig^2_i}+4\delta_{j3}\lambda_{Q_t}^2,\\ {\cal
D}\ln\lambda^2_{L_{\nu j}} &=& -\sum_i{c^{\left(\lambda_L
\right)}_ig^2_i}+3\lambda_{Q_t}^2+
\left(1+4\delta_{j3}\right)\lambda_{\nu_\tau}^2,\\ {\cal D}\ln\lambda^2_{L_{ej}}
&=& -\sum_i{c^{\left(\lambda_L \right)}_ig^2_i}+4\delta_{j3}
\lambda_{\nu_\tau}^2,
\end{eqnarray} where $j$ refers to generation and $i$ refers to the gauge group,
\begin{eqnarray} c^{(\lambda_Q )}=\left( 3,3,{1\over 6},{16\over 3}\right)\, ,\,
c^{(\lambda_L )}=\left( 3,3,{3\over 2},0\right)\, , \end{eqnarray} and we have
used
\begin{eqnarray} {\cal D}\equiv {16\pi^2\over 2}{d\over dt}, \end{eqnarray}
where $t=\ln{(\mu /{\rm GeV})}$ with $\mu$ being the scale.

Now we give the RGEs for the soft SUSY breaking parameters which we need in the
effective
$G_I$ theory. First of all, there are gaugino masses $M_i$ corresponding to each
$g_i$. Secondly, corresponding to each tri-linear superpotential coupling
${\bf \lambda_i}$ there is a tri-linear scalar term with the coupling ${\bf A_i}
{\bf \lambda_i}$ at $M_X$. Finally there are soft scalar mass terms for each of
the the fields
$Q_{L,R}$,
$L_{L,R}$, and $\Phi_{1,2}$. The RGEs for these parameters are as follows:
\begin{eqnarray} {\cal D}M_i&=&b_ig_i^2M_i,\\ {\cal D}A_{Q_{uj}}
&=&\sum_i{c^{\left(\lambda_Q \right)}_ig^2_iM_i}+ \left( 3+4\delta_{j3}
\right)\lambda_{Q_t}^2A_{Q_t}+ \lambda_{\nu_\tau}^2A_{\nu_\tau},\\ {\cal
D}A_{Q_{dj}} &=&\sum_i{c^{\left(\lambda_Q \right)}_ig^2_iM_i}+
4\delta_{j3}\lambda_{Q_t}^2A_{Q_t},\\ {\cal D}A_{L_{\nu j}}
&=&\sum_i{c^{\left(\lambda_L\right)}_ig^2_iM_i}+3\lambda_{Q_t}^2A_{Q_t}+
\left(1+4\delta_{j3}\right)\lambda_{\nu_\tau}^2A_{\nu_\tau},\\ {\cal
D}A_{L_{ej}} &=&\sum_i{c^{\left(\lambda_L\right)}_ig^2_iM_i}+4\delta_{j3}
\lambda_{\nu_\tau}^2A_{\nu_\tau},\\ {\cal
D}M^2_{Q_{jL,R}}&=&-\sum_i{c^{\left(Q_{L,R}\right)}_ig^2_iM_i^2} +
2\lambda_{Q_t}^2X_Q\delta_{j3},\\ {\cal
D}M^2_{L_{jL,R}}&=&-\sum_i{c^{\left(L_{L,R}\right)}_ig^2_iM_i^2} +
2\lambda_{\nu_\tau}^2X_L\delta_{j3},\\ {\cal
D}M^2_{\Phi_1}&=&-\sum_i{c^{\left(\Phi\right)}_ig^2_iM_i^2},\\ {\cal
D}M^2_{\Phi_2}&=&-\sum_i{c^{\left(\Phi\right)}_ig^2_iM_i^2} +
3\lambda_{Q_t}^2X_Q + \lambda_{\nu_\tau}^2X_L , \end{eqnarray} where
\begin{eqnarray} X_Q\equiv M^2_{Q_L}+M^2_{Q_R}+M^2_{\Phi_2}+A^2_{Q_t} ,\\
X_L\equiv M^2_{L_L}+M^2_{L_R}+M^2_{\Phi_2}+A^2_{L_\tau} , \end{eqnarray} and
\begin{eqnarray} c^{(Q_L )}&=&\left( 3,0,{1\over 6},{16\over 3}\right) , c^{(Q_R
)}=\left( 0,3,{1\over 6},{16\over 3}\right),\\ c^{(L_L )}&=&\left( 3,0,{3\over
2},0\right), c^{(L_R )}=\left( 0,3,{3\over 2},0\right),\\ c^{(\Phi )}&=&\left(
3,3,0,0\right), .
\end{eqnarray}

At the scale $M_G$, we assume a universal form to the soft SUSY breaking
parameters i.e. all gaugino masses $M_i(M_G) =m_{1\over 2}$, all tri-linear
scalar couplings $A_i(M_G)= A_0$, and all soft scalar masses $m^2_i
(M_G)=m_0^2$. We also assume
$\lambda_{Q_{t,b}}(M_G)=\lambda_{L_{{\nu_\tau},\tau}}(M_G)$ since quarks and
leptons become unified in SO(10). At the scale $M_I$, we match the $G_I$
effective theory parameters with the MSSM parameters in the usual fashion. We
run all the RGE's according to MSSM \cite{[SUSYYUK],[SUSY],[RGEnote]} down to
the top scale which we take to be 175 GeV. All RGEs are integrated numerically.
We note that since the rank of the SM gauge group is one less than $G_I$, the
soft scalar masses may receive D-term contributions proportional to an
additional parameter at the intermediate scale, however for simplicity we take
this extra unknown parameter to be zero.

We examine three unification scenarios as our examples. Since in all of our
examples no $126+\overline{126}$ representation fields are used to be compatable
with superstring derived models, $SO(10)$ singlets are used to give neutrino
Majorana masses as explained in Ref. \cite{[singlets]}. Also, in all the models
we will discuss the value of the b-quark running mass $m_b$ tends to be in the
vicinity of 4.9 GeV, which is a little high, however thethreshhold corrections
to $m_b$ caneasily be of the order of 10-percent \cite{[raby]}. Also, whatever
operators give masses to the other two generations of down quarks and charged
leptons could be of the right size to fix this problem.

Scenario (a): This model is Case V of Ref. \cite{[MohaLee]}. In the effective
theory above the scale $M_I$, the number $n_H$ of bidoublet, $(2,2,0,1)$, copies
is two and number $n_X$ of SU(2)$_R$ doublet, $(1,2,{1/2},1)+(1,2,{-1/2},1)$
copies belonging to $16+\overline{16}$ representation of SO(10), is four. The
scalar components of these right-handed doublets develop the vacuum expectation
value (VEV) which breaks $G_I$ to the SM gauge group. In this scenerio, we use
$M_I\approx 10^{12}$ GeV and
$M_G\approx 10^{15.6}$ GeV leading to $\alpha_s(M_Z)\approx 0.129$.

Scenario (b): This is the model presented in Ref. \cite{[EMa]}. It is the only
example we use for which D-parity is not broken at $M_G$ and hence left-right
parity ($g_L=g_R$) is preserved in $G_I$. In this model above $M_I$, $n_H=2$,
$n_X=1$ along with one $(2,1,{-1/2},1)+(2,1,{1/2},1)$ superfield belonging to
$16+\overline{16}$ representation as demanded by D-parity, and in addition to
this minimal field content there exist two copies of
$(1,1,{-1/3},3)+(1,1,{1/3},\overline{3})$ and $(1,1,-1,1)+(1,1,1,1)$ from the
$10$ and $120$ representations of SO(10), respectively. This particle content
allows $M_I\sim 1$ TeV with
$M_G\approx 10^{16}$ GeV. We use MSSM below the scale $M_I$ for convenience
although in the original work \cite{[EMa]} the two Higgs doublet model (2HDM)
has been used. The value of
$\alpha_s$ with 2HDM below $M_I$ is about 0.117 and with MSSM it becomes about
0.129. Our result however has very little sensitivity to the value of
$\alpha_s$. One might have expected this scenario to have greater lepton flavor
violation than Scenario (a) has since it has a lower intermediate scale, however
this is not true due to the fact the generationally blind gaugino loop
contribution to the slepton masses is greater in this scenario since $\alpha_G$
is greater.

Scenario (c): This is the model discussed in Ref.\cite{[DKR]}. In this model
$M_G$ is predicted to be exactly the same as in the conventional SUSY SO(10)
breaking with no intermediate scale and the scale $M_I$ can have any value
between the TeV and the GUT scales. In this model, $n_H=1$ and $n_X=3$. Since
there is only one Higgs bidoublet, this model prefers large values of
$\tan{\beta}$ with $\lambda_t =\lambda_b$ at $M_I$. Nevertheless, the
introduction of nonrenormalizable operators can allow for small $\tan\beta$.
Since this model has the unique property that $M_I$ is arbitrary, we use this
model as an example of how the
$\mu\rightarrow e\gamma$ branching ratio changes as a function of $M_I$. The
value of the
$\alpha_s$ at the weak scale is about 0.122.

For all three scenarios, we  run the gauge couplings at two-loops although
we  neglect the small effect with low $\tan\beta$ of the Yukawa couplings on
the gauge coupling running. The $G_I$ gauge beta functions for Scenarios (a) and
(c) are given in Ref. \cite{[DKR]}. The $G_I$ gauge beta functions for Scenario
(b) are given in Ref. \cite{[EMa]}.

The expression with which we calculate the width for $\mu\rightarrow e\gamma$ is
given by Eqs. (29)-(31) in Ref. \cite{[AS]}. This expression \cite{[WidthNote]}
for the width has been determined from the mass interaction basis, which is an
excellent approximation for the low $\tan\beta$ parameter space which we
consider. The CKM elements needed for the amplitude are calculated at the
intermediate scale. We use the neutralino mass matrix as given by Eq. (44) of
Ref. \cite{[SUSY]}.

In Fig. 1-3, we show our results by plotting the function \begin{eqnarray}
l_r\equiv {\rm Log}_{10}{\left( {B\over B_{\rm exp}}\right)}\, , \end{eqnarray} where $B$
is the predicted $\mu\rightarrow e\gamma$ branching ratio and
$B_{\rm exp} =4.9\cdot 10^{-11}$ being the experimental 90 $\%$ confidence limit
upper bound on the branching ratio. In all the figures, we assume that the
universal $M_G$ scale tri-linear soft scalar interaction coupling $A_0 =0$. In
any parameter space which we show, we have checked that the lightest slepton is
not lighter than 43 GeV and the lightest neutralino is not lighter than 20 GeV,
so as to be consistent with their experimental lower bounds of these masses. In
general, we will find that $\mu<0$ gives a greater branching ratio than $\mu
>0$, where $\mu$ is the bi-linear MSSM Higgs superpotential coupling which we
have calculated at the tree level (see, for example, Eq.(22) of Ref.
\cite{[SUSY]}). This is because with
$A_0=0$,
$A_i$ is always negative at the weak scale, and the part proportional to $\mu
\tan\beta+A_i$ in the
$\mu\rightarrow e\gamma$ amplitude has the dominant contribution.

In Fig. 1, we show $l_r$ as predicted by Scenario (a) as a function of the
universal soft mass
$m_0$ for the cases of universal gaugino mass $m_{1/2}=120$ GeV and
$m_{1/2}=200$ GeV . We have taken the $M_G$ scale top Yukawa coupling
$\lambda_{{Q_t}_G} =3.54$. The dashed lines correspond to $\mu <0$ while the
solid ones correspond to $\mu >0$. The upper two lines in the vicinity of
$m_0=150$ GeV correspond to $m_{1/2}=120$ GeV, and the lower two lines in the
same region correspond to $m_{1/2}=200$ GeV. We find the size of lepton flavor
violation through
$\mu\rightarrow e\gamma$ predicted by these two choices of gaugino masses are
fairly typical of those predicted by lighter $ m_{1/2} < 200$ GeV gaugino masses
over the given range of
$m_0$.

In Fig. 2, we show $l_r$ as predicted by Scenario (b) again as a function of the
universal soft mass
$m_0$ for the cases of universal gaugino mass $m_{1/2}=120$ GeV and
$m_{1/2}=200$ GeV . The dashed lines correspond to $\mu <0$ while the solid ones
correspond to $\mu >0$. The upper two lines around $m_0=150$ GeV correspond to
$m_{1/2}=120$ GeV, and the lower two lines in the same region correspond to
$m_{1/2}=200$ GeV. Once again, we have taken the $M_G$ scale top Yukawa coupling
$\lambda_{{Q_t}_G} =3.54$. In this example, we plot only those points where
$l_r$ is less than two orders of magnitude beneath the current experimental
limit.

In Fig. 3, we show $l_r$ in Scenario (c) as a function of $\log_{10}{M_I/GeV}$
for the cases
$\lambda_{{Q_t}_G} =3.54$ and $\lambda_{{Q_t}_G} =1.38$. The dashed lines
correspond to $\mu <0$ while the solid ones correspond to $\mu >0$. The values
of $m_0$ and $m_{1/2}$ are both chosen to be 180 GeV for all the lines. The
upper two lines around $M_I\sim 10^8$ GeV correspond to $\lambda_{{Q_t}_G}
=3.54$, and the lower two lines in the same region correspond to
$\lambda_{{Q_t}_G} =1.38$. Here we want to show the dependence of $l_r$ on the
intermediate scale. For values of
$M_I$ less than about
$10^{6}$ GeV, we see that the two cases of $\lambda_{{Q_t}_G}$ predict
relatively similar values for
$l_r$. Notice that $l_r$ has a maximum for $M_I\sim 10^7$ GeV, rather than $l_r$
montonically increasing as $M_I$ is decreased. This is caused by the fact that
the gaugino contribution to the scalar masses is increased with decreasing $M_I$
and that $\lambda_{L_{\nu_\tau}}$ is no longer of order $\lambda_{Q_t}$ at the
intermediate scale since it has a much lower fixed point than $\lambda_{Q_t}$.

With $M_X$ taken at the reduced Planck scale, as expected one finds an enhanced
branching ratio for $\mu\rightarrow e\gamma$. It is almost impossible to find
parameter space with smaller values of $\mu$ (i.e. $\left| \mu \right| <500$
GeV). As a result the parameter space gets restricted. As an example, in the
scenerio (a) ,the $l_r$ is 1.61 when $m_0=100$ GeV and $m_{1/2}=400$ GeV with
$\mu$ is around 800 GeV. A similar situation also happens in the cases of other
two scenerios.

Between the GUT and the intermediate scale, the large tau-neutrino Yukawa
coupling affects the slepton sector only; the flavor violation in the quark
sector does not get modified significantly by the inclusion of this intermediate
scale. This implies that the parameter space analysis by the constraint from
$b\rightarrow s\gamma$ with the universal soft terms at the GUT breaking
scale
\cite{[bs]} or at the reduced Planck scale
\cite{[we],[us]} still holds in this case unless light SU(2)$_R$ gauge fields
exist \cite{[tgr]}.

In conclusion, we find that an intermediate gauge symmetry breaking is a
significant source of lepton flavor violation in SUSY SO(10) models with GUT
scale uniform soft SUSY breaking terms. In fact, the present limit on
$\mu\rightarrow e\gamma$ rate already puts some limits on the soft breaking
parameters
$m_0$ and $m_{1/2}$. The cause of the lepton flavor violation is simply that the
tau-neutrino's  Yukawa coupling is equal to that of the top quark at the GUT
scale and that the tau neutrino Yukawa coupling reduces the mass of the third
generation sleptons relative to that of the first two generations via the
coupling evolution from the GUT breaking scale down to the intermediate breaking
scale.   Of course in such models, if the scale at which these soft terms appear
is higher than the GUT breaking scale, for example at the reduced Planck scale,
the predicted rate of lepton flavor violation gets enhanced as well. We shall
present other conscequences of an intermediate gauge symmetry, such as
additional
contributions to the electric dipole moments of the electron and neutron, in a
more detailed publication. Although as we have discussed an intermediate scale
can be useful and offers interesting phenomenology, perhaps a difficulty
with the
class of models that we have considered is that there is an additional scale of
symmetry breaking to be explained. However, our results do not depend on the
nature of the mechanism that determines this scale.

We thank K. S. Babu and E. Ma for valuable discussion. This work was supported
by Department of Energy grants DE-FG06-854ER 40224 and DE-FG02-94ER 40837.

\newpage
\leftline{{\Large\bf Figure captions}}
\begin{itemize}

\item[Fig. 1~:] {$l_r\equiv {\rm Log}_{10}{\left( B/B_{\rm exp}\right)}$ in Scenario (a) is
plotted as a function of of the universal soft mass $m_0$.\\ The solid lines
correspond to $\mu >0$ , while the dashed lines correspond to $\mu <0$.\\ The
upper two lines in the vicinity of $m_0=150$ are for $m_{1/2}=120$ GeV, and the
lower two lines are for $m_{1/2}=200$ GeV.\\ $\lambda_{Q_{{t}_G}}=3.54 $ for all
the lines}.
\item[Fig. 2~:] {$l_r$ in Scenario (b) is plotted as a function of the univesal
soft mass $m_{1/2}$,\\ The solid lines correspond to $\mu>0$, and the dashed
lines correspond to $\mu<0$.\\The upper two lines around $m_0=150$ are for
$m_{1/2}=120$ GeV, and the lower two lines in that region are for $m_{1/2}=200$
GeV.\\ $\lambda_{Q_{{t}_G}}=3.54 $ .}
\item[Fig. 3~:] {$l_r$ in Scenario (c) is plotted as a function of
$\log_{10}{M_I/GeV}$.\\ The solid lines correspond to $\mu>0$, the dashed lines
correspond to
$\mu<0$,\\ The upper two lines around $M_I=10^8$ GeV correspond to
$\lambda_{{Q_t}_G} =3.54$, and the lower two lines in the same region correspond
to
$\lambda_{Q_{{t}_G}}=1.38 $. $m_0$ = $m_{1/2}=180$ GeV for all the lines. }

\end{itemize}

\vfill
\begin{figure}[htb]
\centerline{ \DESepsf(figs1.epsf width 12 cm) } \smallskip \nonumber
\end{figure}


\begin{thebibliography}{[001]}
\bibitem{[RGEnote]}One must remember that when $X_Y\equiv \sum_i { Y_i m_i^2}
\neq 0$ at a gauge symmetry breaking scale in which the rank is reduced one must
add an additional term to the soft breaking scalar mass RGEs which scales in a
simple fashion (see, for example, Ref. \cite{[cia]}).
\bibitem{[WidthNote]}(1): To be consistent with our notation, the expression
$\left( I_G^\prime \right)$ in Eq. (30) of  Ref. \cite{[AS]} should be
interperted as $\left( A_e-A_\tau\right)$ with all parameters evaluated at the
scale $M_Z$.\\ (2): We add the weak scale D-term contributions to the soft
breaking masses.
\end{thebibliography}
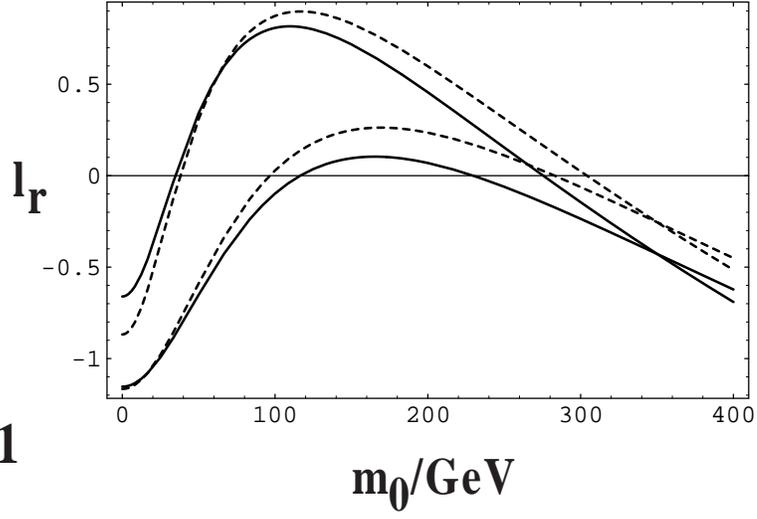
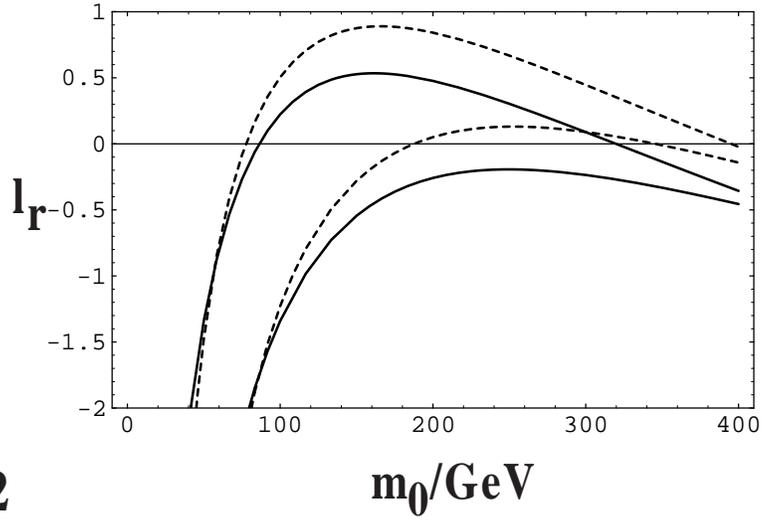
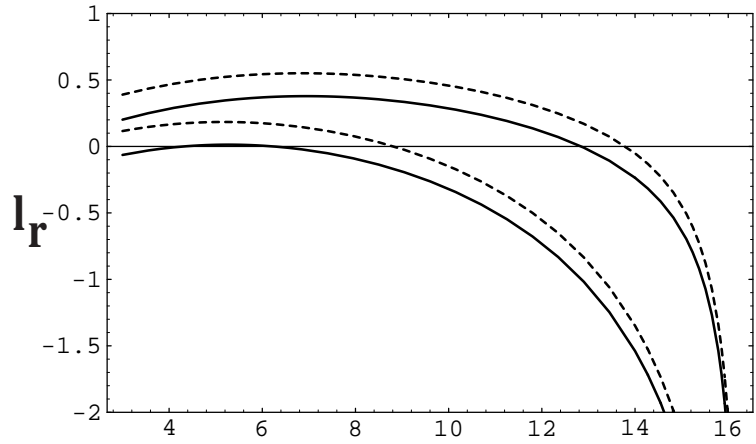
\end{document}